\documentclass[12pt]{article}

\usepackage[round]{natbib}

\usepackage{graphicx}
\usepackage{amssymb,amsfonts,amsmath}

\usepackage{geometry}
\usepackage{setspace}

\newcommand{\thickhline}{%
    \noalign {\ifnum 0=`}\fi \hrule height 1pt
    \futurelet \reserved@a \@xhline
}

\title{Likelihood analysis for a class of beta mixed models}
\author{Wagner Hugo Bonat, 
        Paulo Justiniano {Ribeiro Jr}\footnote{Corresponding author: paulojus@leg.ufpr.br, Dept. Estat\'{\i}stica-UFPR, CP 19.081, Curitiba, PR Brazil, 81.531-990},\\ 
        Walmes Marques Zeviani\\LEG/DEST - Paran\'{a} Federal University}
\date{}

\begin{document}

\maketitle

\begin{abstract}
Beta regression is a suitable choice for modelling continuous response variables taking values on the unity interval. 
Data structures such as hierarchical, 
repeated measures and longitudinal typically induce extra variability and/or dependence
and can be accounted for by the inclusion of random effects.
Statistical inference typically requires
numerical methods, possibly combined with sampling algorithms. 
This class of Beta mixed models is adopted for
the analysis of two real problems with grouped data structures.
We focus on likelihood inference and describe the implemented algorithms.  
The first is a study on the life quality index of industry workers with data 
collected according to an hierarchical sampling scheme.
The second is a study 
assessing the impact of hydroelectric power plants upon
measures of water quality indexes 
up, downstream and at the reservoirs of the dammed rivers,  
with a nested and longitudinal data structure.
Results from different algorithms are reporter for comparison 
including from data-cloning, an alternative to numerical approximations 
which also allows assessing identifiability. 
Confidence intervals based on profiled likelihoods are compared to those obtained by asymptotic quadratic approximations, showing relevant differences for parameters related to the random effects. 
In both cases the scientific hypothesis of interest were 
investigated by comparing alternative models, leading to 
relevant interpretations of the results within each context. 
\end{abstract}

\textit{keywords: hierarchical models, likelihood inference, Laplace approximation, data-cloning, life quality, water quality}


\section{Introduction}
Response variables in the form of proportions, rates and indexes 
are common to different areas such as economics, social and environmental sciences.
They are typically measured in the interval $(0,1)$. This makes
the usual linear (Gaussian) regression model 
inappropriate since it does not ensures predicted values confined 
to the unity domain nor is able capture asymmetries.

Several alternative models are considered in the literature. 
\citet{Kieschnick2003} provides an overview 
and, based on the results of several case studies,  
advocates the adoption of beta regression models.
This is regarded as a flexible class given the diversity
of possible shapes for the distribution function.

Regression models for independent and identically beta 
distributed variables are described by \citet{Paolino2001},
\citet{Cepeda2001}, \citet{Kieschnick2003} and \citet{Ferrari2004}. 
The modelling inherits from the principles of generalised linear models \citep{Nelder1972}, 
relating the expected value of the response variable to covariates through a suitable link function. 
\citet{Cepeda2001}, \citet{Cepeda2005} and \citet{Simas2010} extend the models regressing 
both, mean and dispersion parameters on potential covariates. 
The latter also considers non-linear forms for the predictor.
\citet{smithson:2006} explores the beta regression with an application to IQ data, arguing  
it provides a prudent and meaningful choice compared to alternative approaches.
The beta distribution is able to capture
strongly skewed data, bounded above and below.
It accommodates heterocedasticy, 
and allows for testing hypothesis on location and dispersion separately  
whilst being parsimonious with only two parameters, 
likewise the Gaussian linear model.

Methods for likelihood based inference and model assessment are 
adopted by \citet{Espinheira2008}, \citet{Espinheira2008a} and \citet{Rocha2010}. 
Bias correction for likelihood estimators are developed by \citet{Ospina2006}, 
\citet{ospina_erratum:_2011} and \citet{Simas2010}. 
\citet{Branscum2007} analyses virus genetic distances under the Bayesian paradigma. 
The beta regression is implemented by the \textbf{betareg} package 
\citep{Cribari-neto2010} for the R environment for statistical computing \citep{R}. 
Extended functionality is added for bias correction, 
recursive partitioning and latent finite mixture \citep{grun2011}. 
Mixed and mixture models are discussed by \citet{verkuilen_mixed_2011}.
Time series dependence structure is considered by \citet{McKenzie2007}, 
\citet{Grunwald1993} and \citet{Rocha2010}. 
More recently \citet{Da-Silva2011} adopts a Bayesian beta dynamic model for modelling and 
prediction of time series with an application to the Brazilian unemployment rates.

Dependence structures may arise in other contexts such  
hierarchical model structures, longitudinal and split-plot designs
or any other form of grouping in the sampling mechanism.
Correlation between observations within the same group 
can be induced by inclusion of random effects in the model.
The total variability is therefore decomposed in within and between groups effects. 
As for usual generalised linear mixed models, the
beta mixed models allow for dependent and overdispersed data by 
inclusion of random effects, 
typically assumed to be Gaussian distributed.
Generalised linear mixed models and 
ordinary beta regression models are widely discussed in the literature,
but not beta mixed models, which have recently being considered under 
the Bayesian perspective by \cite{zuniga:2013}. 

Our main goal is to model bounded responses
by beta mixed models, adopting likelihood based methods of inference  
and with discussions in the context of 
the analysis of two real data sets. 
The first is a study on the life quality index of industry workers 
with data grouped on a hierarchical structure.
The second is a comparison of water quality indexes measured 
upstream and downstream hydroelectric power plant reservoirs, 
with data grouped on a longitudinal structure.
The beta mixed model is regarded as a natural choice for both examples. 

Inference require numerical methods since the
likelihood function involves an integral which cannot be solved analytically. 
We consider Gaussian quadrature, quasi Monte Carlo and Laplace 
approximations when integrating the random effects when evaluating the likelihood function. 
We also consider a Markov chain Monte Carlo (MCMC) based algorithm proposed by \cite{Lele2007}
for likelihood inference for generalized linear mixed models which also allows 
investigating identifiability. 
Laplace approximation is less demanding on computing time and suitable for model selection, 
whereas the latter is suitable for further assessment of the best fitted models.
Results are compared with the ones obtained with the computationally less 
demanding linear and non-linear mixed models.


The beta regression model including random effects 
and the adopted methods for likelihood inference are described
 in Section~2.
The two motivating examples are presented in Section~3,
  illustrating the flexibility of the model in accounting for 
  relevant features of the data structures 
  which would be neglected under a standard beta 
regression assuming independent observations. 
The two examples have different justifications and structures for the random effects. 
The first specifies two, possibly correlated, random effects whereas
the second has a nested random effects structure 
as a parsimonious alternative to a fixed effects model.
We compare results obtained with different models and algorithms and 
close with concluding remarks on Section~5.

\section{Beta mixed models}
Let $Y \sim B(\mu,\phi)$ denote a 
beta distributed  random variable with density function 
\begin{equation}
\label{densidadeBeta}
f(y | \mu, \phi) =  \frac{\Gamma(\phi)}{ \Gamma(\mu \phi) \Gamma( (1-\mu)\phi)} y^{\mu \phi - 1} (1 - y)^{(1-\mu)\phi -1}, \quad 0 < y < 1,
\end{equation}
parametrized in terms of mean and dispersion parameters 
\citep{Jorgensen1997,Jorgensen1997b},
where  $\Gamma(.)$ is the Gamma function, $0 < \mu < 1$, 
$\phi > 0$ is a dispersion parameter,   
$E(Y) = \mu$ and  $V(Y) = \frac{\mu(1-\mu)}{(1+\phi)}$. 
For response variables $Y_i \sim B(\mu_i,\phi)$,
the beta regression model \citep{Ferrari2004} 
specifies a linear predictor
$\eta_i = g(\mu_i) = \mathbf{x}_i^T \boldsymbol{\beta} $, 
with $ \boldsymbol{\beta} = (\beta_1, \ldots, \beta_k)^T$ a vector of the $k$ unknown regression coefficients and 
$\mathbf{x}_i = (x_{i1}, \ldots, x_{ik})^T$ a vector of $k$ covariates. 
For the link function $g(\cdot) : (0,1) \to \Re$ we adopt the
\textit{logit} $g(\mu) = \log({\mu / (1-\mu)})$.
Other usual choices are the \textit{probit}, 
\textit{complementary log-log} and \textit{cauchit} \citep{Cribari-neto2010}.

This model does not contemplate possible dependencies such those as induced by
multiple measurements on the same observational unit, over time or space. 
Inclusion of latent random effects on grouped data structure is a parsimonious strategy 
compared to adding parameters to the fixed part of the model
whilst still accounting for nuisance effects.
The random effect model can be specified as follows.
Denote $y_{ij}$ an observation $j = 1, \ldots, n_i$ within group $i = 1, \ldots, q$ and  
$\mathbf{y}_i$ denote a $n_i$-dimensional vector of measurements from the  $i^{th}$ group. 
Let $\mathbf{b}_i$ be a $q$-dimensional vector of random effects and assume  
the responses $Y_{ij}$ conditionally independent with density 
given by (\ref{densidadeBeta}) 
and $g(\mu_{ij}) = \mathbf{x}_{ij}^T \boldsymbol{\beta} + \mathbf{z}_{ij}^T \mathbf{b}_i$,
where  $\mathbf{x}_{ij}$ and $\mathbf{z}_{ij}$
are vector of known covariates with dimensions  $p$ and $q$, respectively, $\boldsymbol{\beta}$ is a $p$-dimensional vector of unknown regression parameters and $\phi$ is the dispersion parameter. 
The model specification is completed assuming Gaussian 
random effects $[\mathbf{b}_i|\Sigma] \sim N(\mathbf{0},\Sigma)$.

\subsection{Parameter estimation}

The model parameters can be estimated by maximising the marginal likelihood 
obtained by integrating the joint distribution $[\mathbf{Y}, \mathbf{b}]$ over the random effects. 
The contribution to the likelihood for the $i^{th}$ group is
\begin{equation*}
\label{verobloco}
f_i(\mathbf{y}_i| \boldsymbol{\beta}, \Sigma, \phi) = \int \prod_{j=1}^{n_i} f_{ij}(y_{ij} | \mathbf{b}_i, \boldsymbol{\beta}, \phi) f(\mathbf{b}_i| \Sigma) d \mathbf{b}_i.
\end{equation*}
Assuming independence among the $q$ groups, 
the full likelihood is given by
\begin{eqnarray}
\label{verocompleta}
L(\boldsymbol{\beta}, \Sigma, \phi) = \prod_{i=1}^q f_i(\mathbf{y}_i | \boldsymbol{\beta}, \Sigma, \phi).
\end{eqnarray}

Evaluation of (\ref{verocompleta}) requires solving the integral $q$ times. 
For the simpler model with a single random effect the integrals are unidimensional. 
More generally, the dimension equals the number of random effects in the model 
which imposes practical limits to numerical methods and approximations 
required to evaluate the likelihood.
 The integrals in our examples have up to five dimensions 
and are solved by Laplace approximation \citep{Tierney1986}
for the results reported here. 
The marginal likelihood is maximised by the algorithm BFGS \citep{Byrd1995} 
as implemented in R.
During our analysis we tried different methods to integrate out the random effects: 
Laplace approximation, Gaussian Quadrature and quasi Monte Carlo.
No differences we detected up to the second decimal place in the  maximised likelihoods. 

We have also considered the sampling based \textit{data cloning} algorithm \citep{Lele2007},  
proposed in the context of maximum likelihood estimation for generalised linear mixed models. 
Data cloning provides tools to assess identifiability \citep{Lele2010} 
which we believe is worth exploring for the beta mixed model. 


The data-cloning algorithm is based on replicating (cloning) $K-times$  
the observations $\mathbf{y}_i$ from each group
generating $N \times K$ cloned data denoted by $\mathbf{y}_i^K$. 
The corresponding likelihood $L^K(\boldsymbol{\beta}, \Sigma, \phi)$  
has the same maximum as (\ref{verocompleta}) 
and Fisher information matrix equals $K$ times the original information matrix. 
The method relies on the Bayesian approach to construct a Monte Carlo Markov chain - MCMC
\citep{Robert2004} algorithm
and using the fact the effect of prior vanishes as the number of clones is increased.
The model is therefore completed by the specification of priors 
$\pi(\boldsymbol{\beta})$, $\pi(\Sigma)$ and $\pi(\phi)$, which
combined with the cloned likelihood, lead to a posterior of the form 
\begin{equation*}
\label{posteriori}
\pi_K(\boldsymbol{\beta}, \Sigma, \phi | y_{ij}) = \frac{ [\int f_i(\mathbf{y}_i | \boldsymbol{\beta}, \Sigma, \phi)f(\mathbf{b}_i | \Sigma) d \mathbf{b}_i ]^K \pi(\boldsymbol{\beta}) \pi(\Sigma) \pi(\phi)} {C(K;y_{ij})}
\end{equation*}
with the normalising constant
\begin{equation*}
C(K;y_{ij}) = \int [ \int f_i(\mathbf{y}_i | \boldsymbol{\beta}, \Sigma, \phi)f(\mathbf{b}_i | \Sigma) d \mathbf{b}_i ]^K \pi(\boldsymbol{\beta}) \pi(\Sigma) \pi(\phi)d \boldsymbol{\beta} d \Sigma d \phi .
\end{equation*}

MCMC algorithms provide a sample from the posterior. By increasing the number $K$ of clones, the posterior mean should converge to the maximum likelihood estimator and 
$K$ times the posterior variance should correspond to 
the asymptotic variance of the MLE \citep{Lele2010}. 
Priors are used to run the algorithm without affecting inference  
as the likelihood can be arbitrarilly weighted by 
increasing the number of clones to the point 
that the effect of priors are negligible. 

Despite the flexibility of the inferential mechanism, 
usual concerns regarding the specification of hierarchical models apply. 
Realistic and suitable models for the problem and available data 
can be complex and need to be balanced against identifiability, 
not often checked nor trivial \citep{Lele2010a}.

Data cloning  provides a straightforward identifiability check 
which can be used for hierarchical models in general. 
\citet{Lele2010} shows that under non-identifiability, the posterior converges to the prior truncated on the non-identifiability space when the number of clones is increased. As a consequence, the largest eigenvector of the parameter's covariance matrix does not converges to zero. More specifically, if identifiable, the posterior variance of a parameter of interest should converge to zero when increasing the number of clones. 

\subsection{Prediction of random effects}

Prediction of random effects are typically required 
as for the examples considered here. 
Under the Bayesian paradigm the predictions can be directly obtained from the posterior distribution of the random effects given by
\begin{equation*}
f_i(\mathbf{b}_i | \mathbf{y}_i, \boldsymbol{\beta}, \Sigma, \phi) = \frac{f_i(\mathbf{y}_i | \mathbf{b}_i, \boldsymbol{\beta}, \phi) f(\mathbf{b}_i | \Sigma)}
                                                                    {\int f_i(\mathbf{y}_i | \mathbf{b}_i, \boldsymbol{\beta}, \phi) f(\mathbf{b}_i | \Sigma) d \mathbf{b}_i} ,
\end{equation*}
which does not have a closed expression for the beta model. 
 The posterior mode maximizes $f_i(\mathbf{y}_i | \mathbf{b}_i, \boldsymbol{\beta}, \phi) f(\mathbf{b}_i | \Sigma)$
 providing a point predictor for $\hat{\mathbf{b}}_i$ and
 empirical Bayes predictions can be obtained by replacing the unknown parameters by their maximum likelihood estimates.

\section{Examples}

\subsection{Income and life quality of Brazilian industry workers}

The Brazilian industry sector \textit{worker's life quality index} (IQVT, acronym in Portuguese) combines 25 indicators from eight thematic areas: housing, health, education, integral health and workplace safety, skill development, work attributed value, corporate social responsibility, participation and performance stimulus. The index is constructed following the same premisses of the united nations human development index\footnote{http://hdr.undp.org/en/humandev/}. 
Values are expressed in the unity interval and the closer to one, the 
higher the industry's worker life quality.

A pool was conducted by the Industry Social Service\footnote{Servi\c{c}o Social da Ind\'ustria - SESI} in order to assess worker's life quality in the Brazilian industries. The survey included $365$ companies from 
nine out of the $27$ 
Brazilian federative units. 
IQVT  was computed for each company from questionnaires applied to workers according to a sampling design. Companies provided additional information on 
budget for social benefits and other quality of life related initiatives.

\begin{figure}[htbp]
\centering
\includegraphics{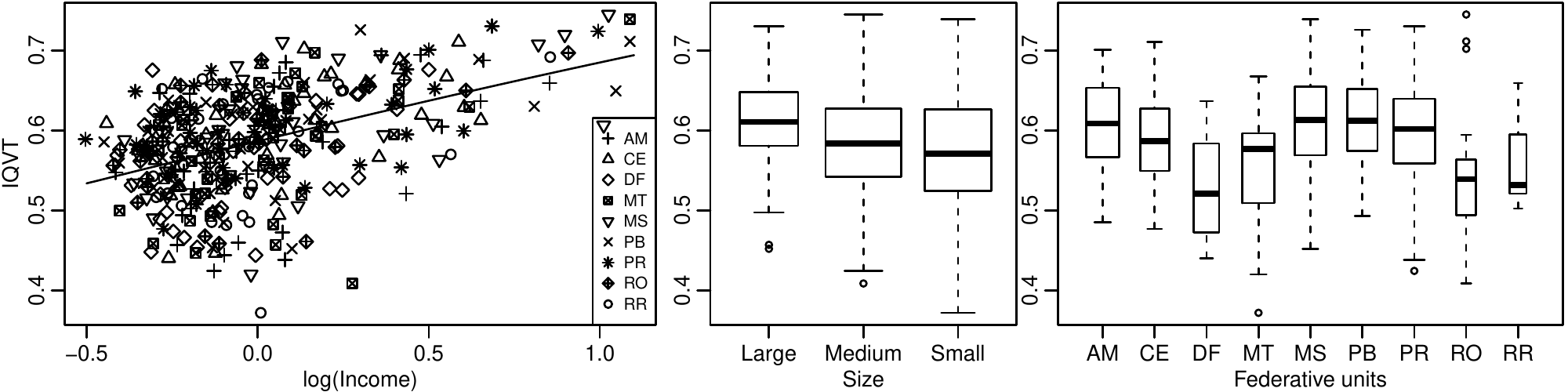}
\caption{IQVT related to (centred log) average income, company size and federative unit.}
\label{fig:descritivaIQVT}
\end{figure}

A suitable model is aimed to assess the effects on IQVT of two company related covariates, average namely \textit{income} and \textit{size}. 
The first is simply the total of salaries divided by the number of workers 
expressing the capacity to fulfil individual basic needs such as food, 
health, housing and education. 
The second reflects the industry's quality of life management capability. 
There is a particular interest in learning whether larger companies  
with 500 or so workers, typically multinational, 
working under regimes of worldwide competition,
provide better life standards in comparison with medium (100 to 499 workers) 
and small (20 to 99 workers) sized industries.
The federative unit where the company is located is expected to be influential
due to varying local legislations, taxing and further economic and political conditions. 
Plots on Figure~\ref{fig:descritivaIQVT} suggests IQVT is associated with 
income, size and location. 
The income is expressed in logarithmic scale centred around their average.

\setkeys{Gin}{width=0.99\textwidth}

The beta random effects model for IQVT is
\begin{align*}
  Y_{ij} | \mathbf{b}_i &\sim B(\mu_{ij} , \phi) \\
  g(\mu_{ij}) &= (\beta_0 + b_{i1}) + \beta_1 Medium_{ij} + \beta_2 Small_{ij} + (\beta_3 + b_{i2}) Income_{ij} \\
  \mathbf{b}_i &\sim NMV(\mathbf{0}, \Sigma)  \mbox{  with  } \Sigma = \left[\begin{array}{rr} 
1/ \tau_1^2 &\rho\\
\rho& 1 / \tau_2^2 \\
\end{array}\right] ,
\end{align*}
parametrized such that $\beta_0$ is associated with large sized companies and 
$\beta_1$ and $\beta_2$ are differences for the medium and small sized companies, 
respectively. 
A random intercept $b_{i1}$ and slope $b_{i2}$ associated with \textit{income}
account for the effect of the federative units. 
Model parameters to be estimated consist of 
the regression coefficients $(\beta_0, \beta_1, \beta_2, \beta_3)$, 
the random effects covariance parameters $(\tau_1^2, \tau_2^2, \rho)$    
and the precision parameter $\phi$. 

A sequence of sub-models are defined for testing relevant effects. 
Model~1 is the null model with simply the intercept. 
Model~2 includes the covariate \textit{size} and Model~3 the \textit{income}. 
Model~4 adds random intercepts and Model~5
adds a random slope to \textit{income}.
For comparison, we also fit the  corresponding linear mixed Gaussian  (LMM) and non-linear logistic models (NLMM), 
widely used in practice.



Parameter estimates for the beta models using Laplace approximation for the random effects are given 
in the top part of Table~\ref{tab:comparativoIQVT}
and maximised log-likelihoods for the five model structures are given in Table~\ref{tab:comparativoIQVT}
along with the ones for the linear and non-linear models.

 \begin{table}[] 
 \centering 
 \caption{Parameter estimates for the beta models (top) and maximised likelihood for the alternative models (bottom) - IQVT.}  
 \label{tab:comparativoIQVT} 
 \begin{tabular}{lccccc} 
   \hline 
 Parameter & Model 1 & Model 2 & Model 3 & Model 4 & Model 5 \\ 
   \hline 
 $\beta_0$ & 0.35 & 0.45 & 0.43 & 0.40 & 0.40 \\  
   $\beta_1$ &  & -0.11 & -0.09 & -0.07 & -0.07 \\  
   $\beta_2$ &  & -0.16 & -0.14 & -0.13 & -0.13 \\  
   $\beta_3$ &  &  & 0.42 & 0.47 & 0.47 \\  
   $\phi$ & 53.97 & 56.80 & 72.86 & 94.19 & 94.19 \\  
   $\tau_1^2$ &  &  &  & 62.36 & 62.35 \\  
   $\tau_2^2$ &  &  &  &  & 51480.17 \\  
   $\rho$ &  &  &  &  & 0.85 \\  
 \hline Model &  \multicolumn{5}{c}{Maximised likelihood} \\ 
 Beta & 472.20 & 481.51 & 526.94 & 561.79 & 561.80 \\  
   LMM & 470.42 & 479.96 & 523.85 & 558.89 & 558.90 \\  
   NLMM & 470.42 & 479.96 & 523.77 & 558.96 & 558.96 \\  
    \hline 
 \end{tabular} 
 \end{table} 
Comparison of models 1-3 confirms the relevance of the covariates with 
estimates of the precision parameter $\phi$ increasing 
from $53.97$ on Model~1 to $72.85$ on Model~3.
The random intercept added in Model~4  
clearly further improves the fit 
expressing the variability of the IQVT among the federative units 
with an increase of $34.85$ in the log-likelihood.  
Addition of the random slope did not prove relevant. 
Model~4 including the two covariates and just the random intercept is therefore the model of choice.

The beta mixed model is not commonly adopted in the literature 
and this motivates us to consider the data cloning as a distinct    
approach for likelihood computations which also allows for assessing the model identifiability. 
The results are reassuring with similar estimates and standard errors 
obtained by maximization of the numerically integrated marginal likelihood and data cloning  
as shown in Table~\ref{tab:margcloneIQVT}.

 \begin{table}[] 
 \centering 
 \caption{Parameter estimates and standard errors for Model 4 by marginal likelihood and data-cloning - IQVT.}  
 \label{tab:margcloneIQVT} 
 \begin{tabular}{lcccc} 
 \hline Parameter & \multicolumn{2}{c}{Marginal likelihood} & \multicolumn{2}{c}{Data-clone}\tabularnewline 
  & Estimate & Std. error & Estimate & Std. error \\  
   \hline 
 $\beta_0$ & 0.40 & 0.05 & 0.40 & 0.05 \\  
   $\beta_1$ & -0.07 & 0.03 & -0.07 & 0.03 \\  
   $\beta_2$ & -0.13 & 0.03 & -0.13 & 0.03 \\  
   $\beta_3$ & 0.47 & 0.04 & 0.47 & 0.04 \\  
   $\phi$ & 94.19 & 7.03 & 94.17 & 6.98 \\  
   $\tau_1^2$ & 62.36 & 32.00 & 62.03 & 32.08 \\  
    \hline 
 \end{tabular} 
 \end{table} 
Interval estimates obtained by both,
the asymptotic quadratic approximation with standard errors returned by data cloning
and by profiling the marginal likelihoods are presented in Table~\ref{tab:icIQVT}.
The latter can be asymmetric and with coverages 
closer to nominal values. 
Intervals are similar for all the parameters 
except for $\tau_1^2$ with an artefactual negative lower bound for the quadratic approximation.

\begin{figure}[htbp]
\centering
\includegraphics{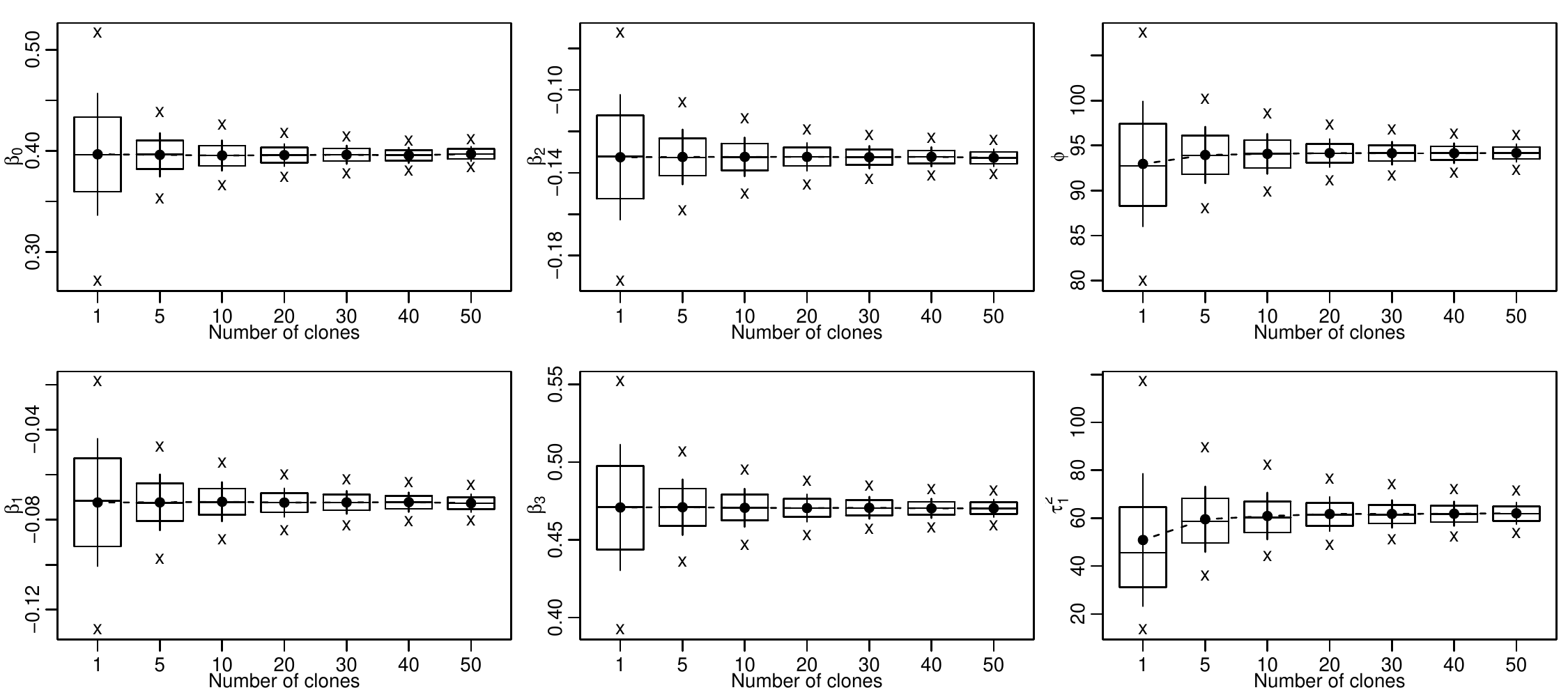}
\caption{Sampled parameter values for different number of clones, Model 4 - IQVT.}
\label{fig:boxplot}
\end{figure}

 \begin{table}[] 
 \centering 
 \caption{Asymptotic and profile likelihood based confidence intervals, Model 4 - IQVT.}  
 \label{tab:icIQVT} 
 \begin{tabular}{lcccc} 
 \hline Parameter & \multicolumn{2}{c}{Asymptotic} & \multicolumn{2}{c}{Profile}\tabularnewline 
  & 2.5\% & 97.5\% & 2.5\% & 97.5\% \\  
   \hline 
 $\beta_0$ & 0.30 & 0.50 & 0.29 & 0.50 \\  
   $\beta_1$ & -0.13 & -0.02 & -0.13 & -0.02 \\  
   $\beta_2$ & -0.19 & -0.07 & -0.19 & -0.07 \\  
   $\beta_3$ & 0.39 & 0.55 & 0.39 & 0.55 \\  
   $\phi$ & 80.49 & 107.84 & 81.09 & 108.65 \\  
   $\tau_1^2$ & -0.85 & 124.91 & 19.74 & 156.48 \\  
    \hline 
 \end{tabular} 
 \end{table} 
Identifiability can be assessed by the data cloning method as described in Section~2. 
We use the package $dclone$ \citep{dclone2010}, with the JAGS \citep{Plummer03jags:a} MCMC engine with $1$, $5$, $10$, $20$, $30$, $40$ and $50$ clones. For each number of clones we use $3$ independent chains of size $6500$,  and burn-in of $1500$. Results are summarised in Figure~\ref{fig:boxplot} with chains increasingly concentrated around the maximum likelihood estimate with increasing number of clones.

\setkeys{Gin}{width=0.95\textwidth}

Figure~\ref{fig:boxplot} also allows comparison between results for
usual Bayesian inference ($K=1$) with for the $K$ fold cloned data. 
The adopted flat normal prior (zero mean and precision $0.001$) 
for the regression parameters was not influential whereas   
the Gamma$(1,0.001)$ prior for $\phi$ and $\tau_1^2$ 
showed different results for $K=1$ and $K=50$. 

\begin{figure}[htbp]
\centering
\includegraphics{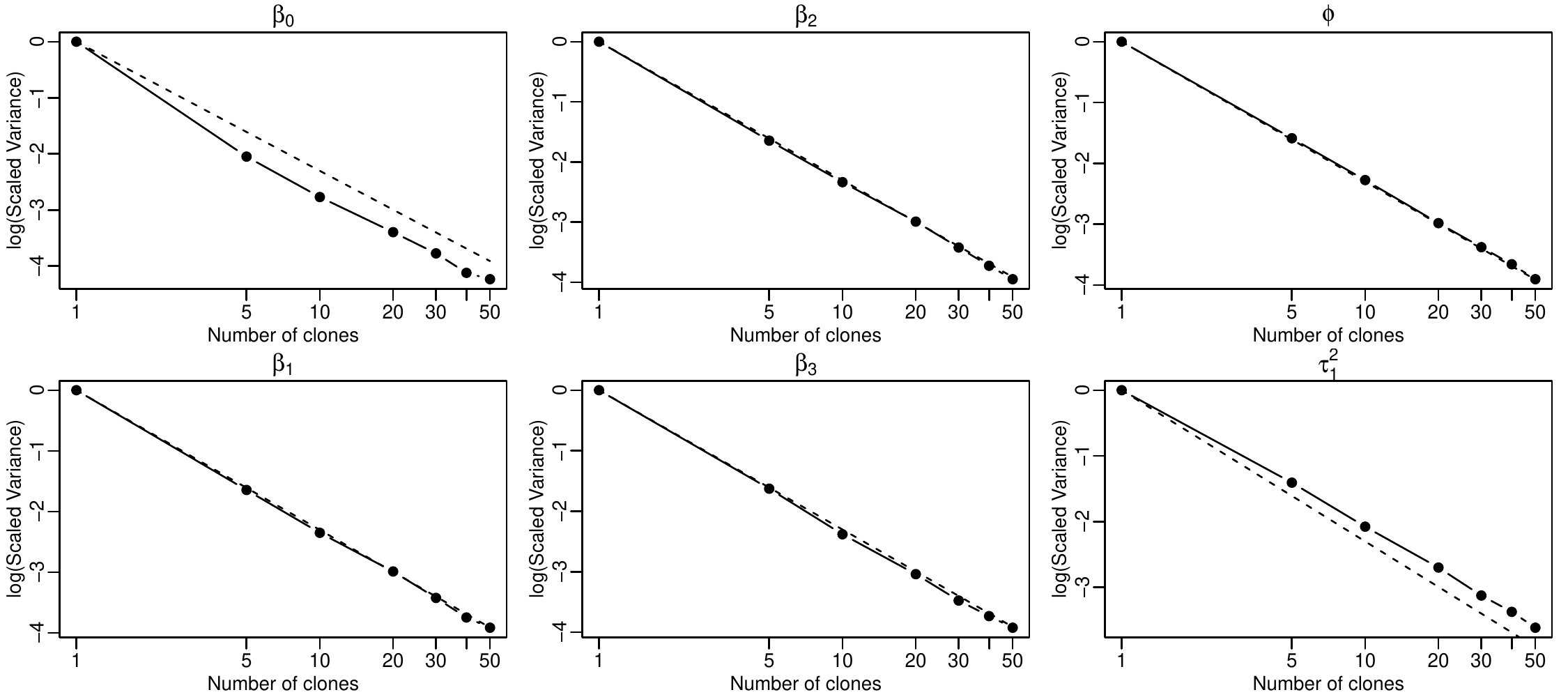}
\caption{Identificability diagnostics with data-cloning for the beta model with random intercept - IQVT.}
\label{fig:estimabilidadeIQVT}
\end{figure}

Under identifiability, the posterior variance should converge to zero for increasing number of clones $K$ with variance decreasing at rates $1/K$. Such trend is detected as shown in Figure~\ref{fig:estimabilidadeIQVT} 
which uses logarithmic scales to ease the visualisation. 
Variances decrease satisfactorily at nearly expected rates with a slight but not relevant difference for the $\tau_1^2$ parameter supporting the conclusion that Model~4 is identifiable for the current data.

\setkeys{Gin}{width=0.95\textwidth}

Fitted coefficients support the initial conjectures that the size has a relevant effect on the IQVT with expected decrease of $3.01\%$ and $5.70\%$ changing from large to medium and small sizes, respectively. These predictions are obtained setting the other factors to baseline and/or zero values. Increasing income clearly positively affects the IQVT, confirming and quantifying the expected behaviour. The fitted random intercepts confirm the existence of a substantial variation 
in the quality of life among the federative units. 
Table~\ref{tab:preditos} summarises IQVT predicted 
for different federative units and company sizes and computed 
for lower (R$\$500.00$) and higher (R$\$2,500.00$) levels of income.

\begin{table}
\caption{\label{tab:preditos} Predicted indexes and percent differences (within parenthesis) to the global average for Model~4 - IQVT.}
\center
\begin{tabular}{cccc|ccc}
\hline 
Fed. & \multicolumn{3}{c|}{R\$ 500,00} & \multicolumn{3}{c}{R\$ 2.500,00}\tabularnewline
Unity  & Large & Medium & Small & Large & Medium & Small\tabularnewline
\hline 
AM & 52.91(1.52) & 51.11(1.58) & 49.60(1.63) & 70.55(0.95) & 69.02(1.00) & 67.72(1.04)\tabularnewline
CE & 54.48(4.52) & 52.68(4.70) & 51.17(4.85) & 71.84(2.80) & 70.35(2.95) & 69.08(3.07)\tabularnewline
DF & 46.5(-10.77) & 44.71(-11.13) & 43.23(-11.43) & 64.95(-7.06) & 63.29(-7.39) & 61.88(-7.68)\tabularnewline
MT & 50.82(-2.49) & 49.01(-2.58) & 47.51(-2.65) & 68.78(-1.58) & 67.21(-1.66) & 65.87(-1.73)\tabularnewline
MS & 54.22(4.04) & 52.42(4.20) & 50.92(4.33) & 71.63(2.51) & 70.14(2.64) & 68.86(2.75)\tabularnewline
PB & 56.91(9.20) & 55.13(9.58) & 53.64(9.90) & 73.79(5.60) & 72.37(5.90) & 71.15(6.16)\tabularnewline
PR & 53.83(3.29) & 52.03(3.42) & 50.52(3.52) & 71.31(2.04) & 69.81(2.15) & 68.52(2.24)\tabularnewline
RO & 49.17(-5.66) & 47.36(-5.86) & 45.86(-6.03) & 67.34(-3.64) & 65.73(-3.82) & 64.36(-3.97)\tabularnewline
RR & 50.11(-3.85) & 48.31(-3.99) & 46.80(-4.1) & 68.17(-2.45) & 66.58(-2.58) & 65.22(-2.68)\tabularnewline
\hline 
\end{tabular}
\end{table}

Table~\ref{tab:preditos} shows positive effects for
Mato Grosso do Sul (MS), Paran\'{a} (PR), Amazonas (AM), Cear\'{a} (CE)
and Para\'{\i}ba (PB) with the latter being the best case where IQVT was $9.9\%$ above the global average for small size business with average income of $R\$500.00$. 
Negative effects were estimated for Mato Grosso (MT), Roraima (RR), Rond\^{o}nia (RO) and 
Distrito Federal (DF) being the worse case with IQVT $11.43\%$ below the global average.

\begin{figure}[htbp]
\centering
\includegraphics{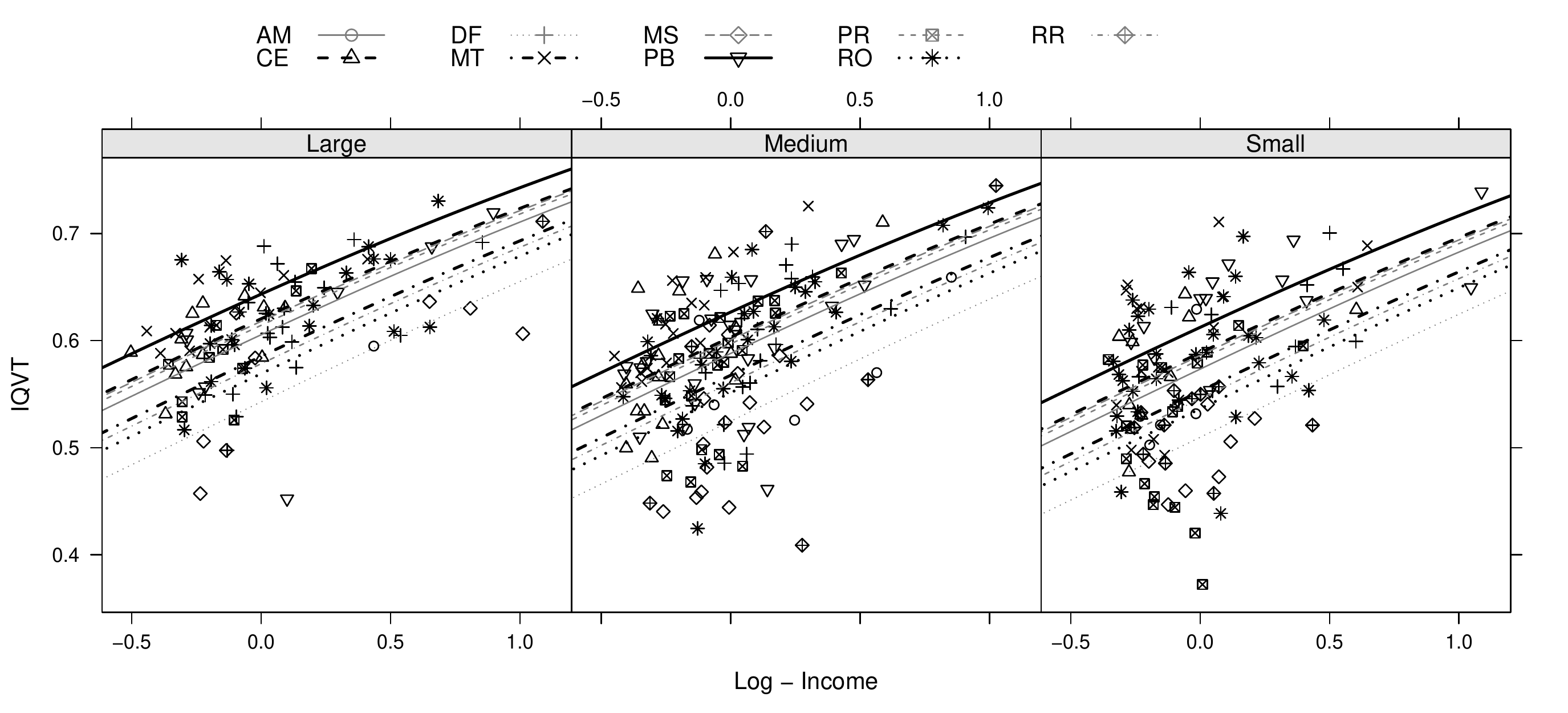}
\caption{Observed and fitted values for the random intercept model for each business size.}
\label{fig:ajusteIQVT}
\end{figure}

The differences for incomes around R$\$500.00$ 
become smaller for incomes around R$\$2,500.00$, 
indicating a decreasing influence of company size and federative unity for increasing incomes. 
The more pronounced effect 
for low incomes are compatible with Brazilian conditions.  There are several governmental supporting policies 
which effectively improve quality of life for low income workers such as the  
social assistance unified system, the young agents programme, 
social and food security, food support, popular restaurants, community catering, 
family health, maintenance allowance, development educational fund among other
Brazilian governmental social programs\footnote{listed at http://www.portaltransparencia.gov.br}. 
Additionally companies internal supporting incentives for low income workers such as catering, transportation, 
basic shopping supply,
among others, make the workplace relevant for the worker quality of life.
On the other hand, the higher the income, the lesser the dependence on such benefits 
and income becomes the main, if not the single, maintainer of life quality and therefore less influenced 
by conditions such as those reflected by industry size and federative unit. Interpretations based on the fitted model are therefore compatible with the subjective information about the working circumstances in the country. The observed data and fitted values for the random intercept model for each business size is shown on Figure~\ref{fig:ajusteIQVT}. 

\setkeys{Gin}{width=0.99\textwidth}

Figure~\ref{fig:ajusteIQVT} shows IQVT values concentrated between $0.35$ and $0.80$ and within this range the relation with the log-income is nearly linear with a satisfactory adherence to the data. Apparent outliers did not show influence on the overall model fit.

\subsection{Water quality on power plant reservoirs}

The energy company COPEL operates 16 hydroelectric power plants in Paran\'{a} State, Brazil, generating over 4.500 MW. 
The reservoir lakes 
are also used for leisure activities, navigation and water supply. 
Effective functioning of the power plants depends on the
quality of the water, which affects the growth of organisms and aquatic flora.
Assessing possible impacts of the reservoirs on water quality 
is relevant for the water supply and environmental hazards. 
In compliance with operating licenses, 
the concessionaire company regularly monitors the water quality  
upstream, downstream and at the reservoirs of the dammed rivers. 

Monitoring is based on the comparison of nine water quality indicators against reference 
values given by standards for water supply. 
The water quality indicators are: dissolved oxygen, temperature, faecal coliform, water pH, biochemical oxygen demand (DBO), total nitrogen, total phosphorus, turbidity and total solids. 
The indicators are also combined to produce a single value of a water quality index 
(IQA, acronym in Portuguese) 
based upon a study conducted in the 70's 
by the US National Sanitation Foundation
and adapted by the Brazilian company CETESB\footnote{Companhia de Tecnologia de Saneamento Ambiental}. 

Monitoring aims to detect changes in the water quality, 
possibly attributable to the presence of the dams. 
Water quality measurements taken at locations considered directly affected and unaffected by the reservoir are compared. 
More specifically, 
measurements taken upstream the main river are considered unimpacted reference values 
whereas measurements taken at the reservoir and downstream are considered 
potentially affected by the water contention and passage throughout the power plant.

Water quality indicators are measured quarterly on 
the 16 operating hydroelectric power plants and we
consider the data collected during 2004. 
The main interest is the covariate \textit{Local}, 
with levels \textit{upstream}, \textit{reservoir} and \textit{downstream}
controlled for effects of the \textit{power plant} and 
the \textit{quarter} of data collection.
This amounts to 190 data with 12 measurements (4 quarters $\times$ 3 locations) 
for each of the 16 power plants with only two missing data.  

\setkeys{Gin}{width=0.95\textwidth}
\begin{figure}[htbp]
\centering
\includegraphics{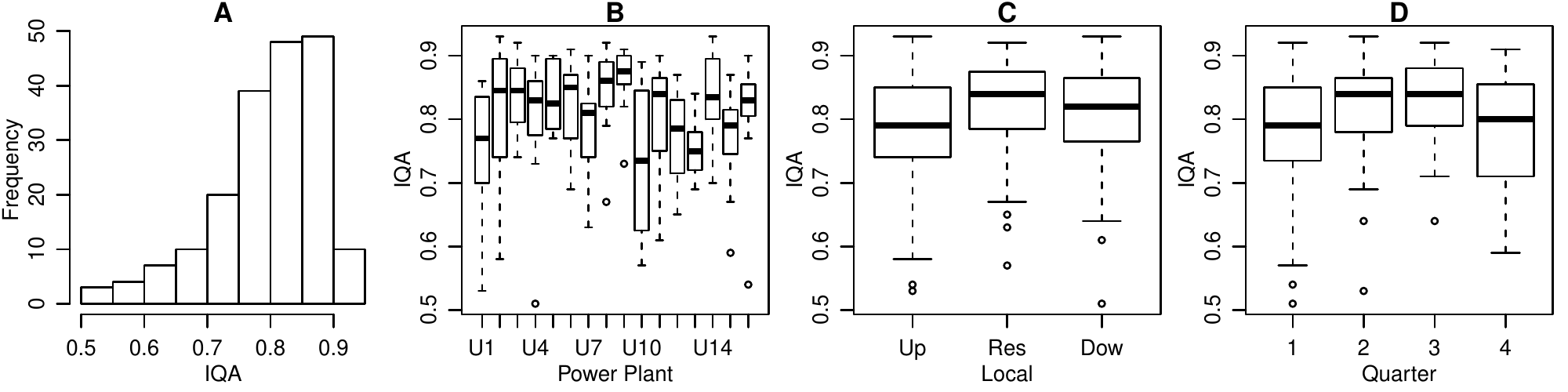}
\caption{Summaries for the IQA data.}
\label{fig:descritiva}
\end{figure}

The left asymmetry in Figure~\ref{fig:descritiva}-A 
is typical for this kind of data. 
IQA varies between power plants as seen in \ref{fig:descritiva}-B. 
Figure~\ref{fig:descritiva}-C suggests an 
increase from upstream to the reservoir and a decrease from reservoir to downstream. 
Figure~(\ref{fig:descritiva}-D) shows lower values for 
the first and forth quarters, the warmer periods,  
a pattern expected to be repeated over the years.

This brief exploratory analysis suggests that in order to investigate 
the effect of the position relative to the dam (\textit{Local}), 
the effects of quarter and power plant should be accounted for, 
possibility with distinct quarter effects for different plants.
Main effects and interactions would amount for 80 degrees of freedom
under a  fixed effects model. 
Instead, we regard the power plants as a sample from a population of environments
and add a corresponding random effect term in the model.  
This is a convenient assumption for our intended method of analysis 
and proved sound for this particular application.

IQA on the the $i^{th}$ relative location, $j^{th}$ power plant e $t^{th}$ quarter 
is modelled by
\begin{align*}
  Y_{ijt} | b_j, b_{jt} &\sim B(\mu_{ijt}, \phi) \\
  g(\mu_{ijt}) &= \beta_0 + \beta_{1i} + \beta_{2t} + b_{j} + b_{jt} \\
  b_j &\sim N(0,\tau_U^2) \;\;;\;\; b_{jt} \sim N(0,\tau_{UT}^2) , 
\end{align*}
Under the adopted parametrization, 
$\beta_{1i}$,  $i= 2,3$ quantifies the changes from upstream to reservoir and downstream, respectively. 
Likewise $\beta_{2t}$, $t=2,3,4$ are differences between the first quarter and the others. 
The random intercept $b_j$ captures the deviations of each power plant to the overall mean and  
$b_{jt}$ are the random effects of each quarter within each power plant. 

Hypotheses of interest are tested comparing submodels defined by 
adding terms $\beta_{1i}$, $\beta_{2j}$, $b_{j}$ and $b_{jt}$ 
sequentially up to the full model. 
Parameter estimates are presented in Table~\ref{tab:ajuste}.  
Numerical estimates are obtained by the BFGS algorithm for maximizing the 
marginal likelihood with Laplace approximation for the numerical integration of the random effects. 
The difference of only $1.1091$ between Model~5 and~6
indicates unnecessary the inclusion of $b_{jt}$.

 \begin{table}[] 
 \centering 
 \caption{Parameter estimates for the beta models (top) and maximised likelihood for the alternative models (bottom) - IQA.}  
 \label{tab:ajuste} 
 \begin{tabular}{lcccccc} 
   \hline 
 Parameter & Model 1 & Model 2 & Model 3 & Model 4 & Model 5 & Model 6 \\ 
   \hline 
 $\beta_0$ & 1.40 & 1.27 & 1.14 & 1.14 & 1.15 & 1.15 \\  
   $\beta_{12}$ &  & 0.23 & 0.23 & 0.24 & 0.24 & 0.24 \\  
   $\beta_{13}$ &  & 0.15 & 0.15 & 0.16 & 0.15 & 0.16 \\  
   $\beta_{22}$ &  &  & 0.21 & 0.22 & 0.22 & 0.22 \\  
   $\beta_{23}$ &  &  & 0.29 & 0.31 & 0.32 & 0.32 \\  
   $\beta_{24}$ &  &  & 0.05 & 0.05 & 0.06 & 0.06 \\  
   $\phi$ & 23.36 & 24.25 & 25.78 & 30.47 & 42.19 & 42.20 \\  
   $\tau_U^2$ &  &  &  & 28.97 &  & 43.54 \\  
   $\tau_{UT}^2$ &  &  &  &  & 11.19 & 15.04 \\  
 \hline Model &  \multicolumn{6}{c}{Maximised likelihood} \\ 
 Beta & 215.38 & 218.90 & 224.62 & 231.04 & 237.08 & 238.19 \\  
   LMM & 198.23 & 202.12 & 208.68 & 213.68 & 220.39 & 225.01 \\  
   NLMM & 198.23 & 202.12 & 208.72 & 214.88 & 223.12 & 223.91 \\  
    \hline 
 \end{tabular} 
 \end{table} 

The likelihood evaluation 
for the largest model requires numerical approximation
of a five dimensional integral, for each reservoir.
Dimensionality of the integrals greatly increases the computational burden 
for the algorithms based on Gauss-Hermite, Monte-Carlo integration or Laplace approximation.
The alternative data-cloning algorithm does not 
demand integral approximation nor numerical maximization
and the computational burden is determined by the sampling strategy and fits for increasing number of clones. 
Table~\ref{tab:comparativo} presents the parameter estimates for Model~5 obtained by both ways, 
integration by Laplace approximation and data-cloning.
The regression coefficients are similar however with differences on the standard errors. 
Overall, smaller values were obtained with the numerically integrated marginal likelihood
which demanded more accurately recomputing of the numerical Hessian for the fitted model.

 \begin{table}[] 
 \centering 
 \caption{Parameter estimates and standard errors for Model~5  by marginal likelihood and data-cloning - IQA.}  
 \label{tab:comparativo} 
 \begin{tabular}{lcccc} 
 \hline Parameter & \multicolumn{2}{c}{Marginal likelihood} & \multicolumn{2}{c}{Data-clone}\tabularnewline 
  & Estimate & Std. error & Estimate & Std. error \\  
   \hline 
 $\beta_0$ & 1.15 & 0.09 & 1.15 & 0.10 \\  
   $\beta_{12}$ & 0.24 & 0.05 & 0.24 & 0.07 \\  
   $\beta_{13}$ & 0.15 & 0.01 & 0.15 & 0.07 \\  
   $\beta_{22}$ & 0.22 & 0.01 & 0.22 & 0.13 \\  
   $\beta_{23}$ & 0.32 & 0.03 & 0.31 & 0.13 \\  
   $\beta_{24}$ & 0.06 & 0.01 & 0.06 & 0.13 \\  
   $\phi$ & 42.19 & 4.14 & 42.30 & 5.32 \\  
   $\tau_{UT}^2$ & 11.19 & 3.31 & 10.99 & 3.12 \\  
    \hline 
 \end{tabular} 
 \end{table} 

\begin{figure}[tbhp]
\centering
\includegraphics{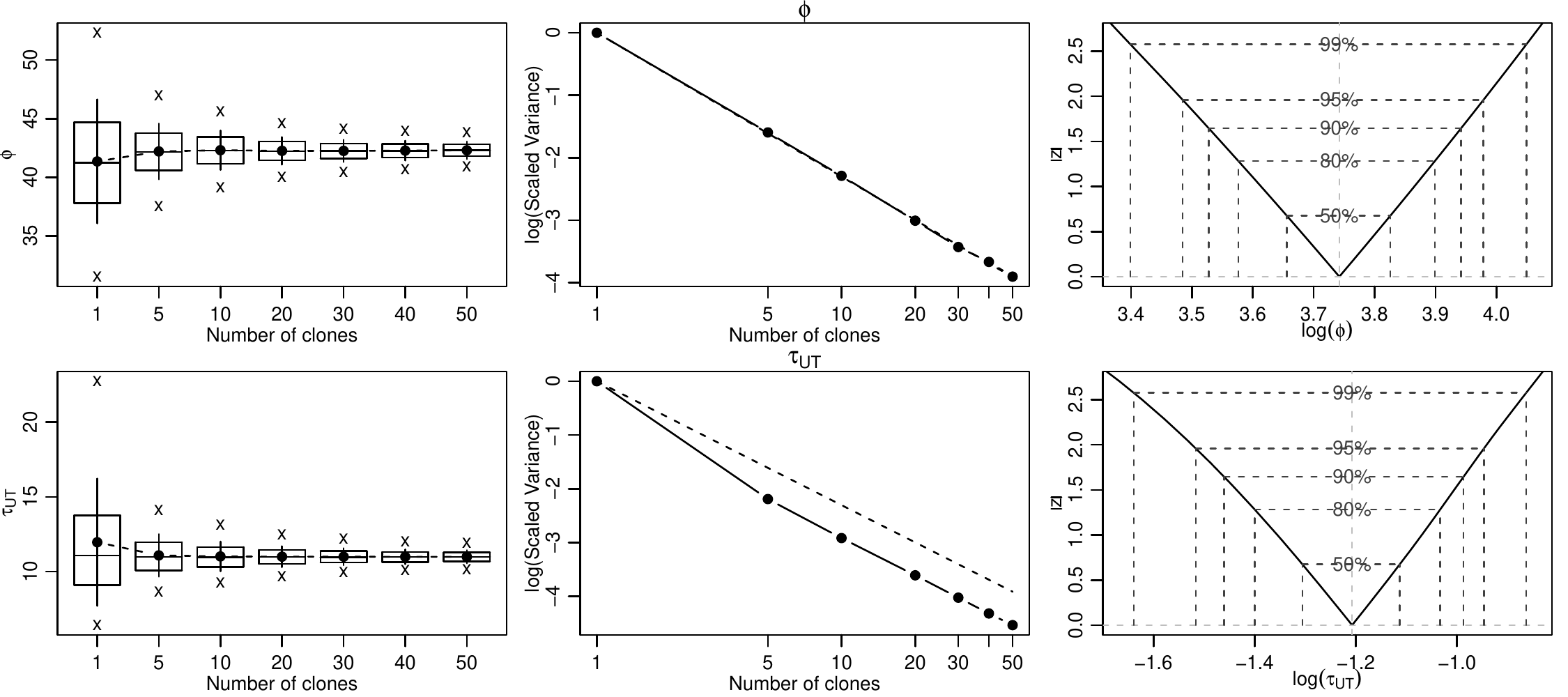}
\caption{Identifiability diagnostic plots and profile likelihoods for precision parameters under Model~5 - IQA.}
\label{fig:estimabilidadeIQA}
\end{figure}

Quadratic approximation of the likelihood does not hold and symmetric confidence intervals based on the standard deviations are clearly inappropriate for parameters 
$\phi$ e $\tau_{UT}$ for which we compute intervals based on profile likelihoods.
Right hand panels of Figure~\ref{fig:estimabilidadeIQA} show the profile likelihoods for the precision parameters 
reparametrised on the logarithmic scale for computational convenience. 
Left hand and middle panels are data-cloning identifiability diagnostic plots.
The profile likelihood for $\log(\tau_{UT})$ 
is asymmetric with this parameter being more sensitive
than $\phi$ to the choice of prior, 
as indicated by the comparison of boxplot for original ($K=1$) and cloned data. 
The scaled variance plots for $\log(\tau_{UT})$ shows a 
slightly faster than expected decay in variance for the corresponding number of clones, 
however the larger eigenvalue for the covariance matrix were found to be
always smaller than 1.1, an indicator of identifiability.

\setkeys{Gin}{width=0.99\textwidth}
 
\begin{figure}[tbhp]
\centering
\includegraphics{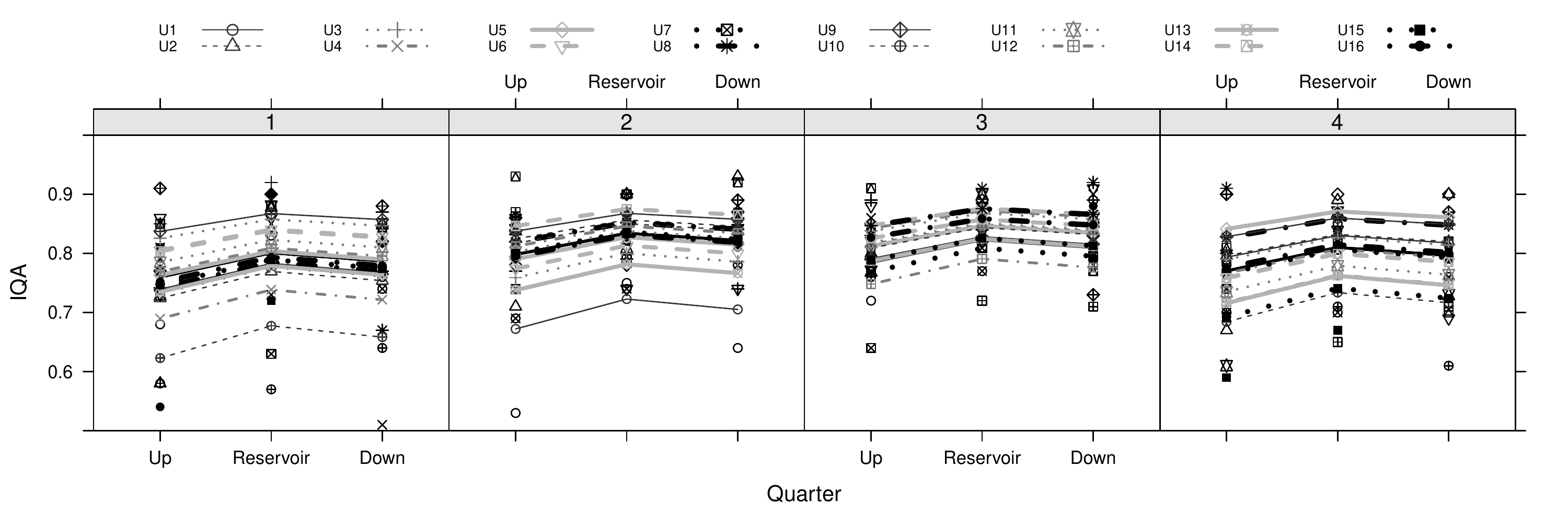}
\caption{Predicted values under Model~5 connected by lines - IQA.}
\label{fig:ajusteIQA}
\end{figure}

Empirical Bayes predictions of the random effects are connected by lines in
Figure~\ref{fig:ajusteIQA}.
Setting random effects to zero, 
the fitted model predicts that the IQA increases $5.39\%$ from upstream to the reservoir and $3.55\%$ from up to downstream. 
The analysis confirms lower IQA values for the warmer temperatures first and forth quarters compared with the milder temperatures second and third quarters. 
This is expected to be a cyclic behaviour over the years. 
Significance of random effects implies that the differences vary between power plants and quarters. 
In summary, the overall pattern is that the IQA substantially improves from upstream to the reservoir 
however shifting back closer to the original values downstream, however with substantial variation of the differences between the power plants.

\setkeys{Gin}{width=0.99\textwidth}

The adopted model and the algorithms implementing inferential methods proved satisfactory. 
Some extreme measurements taken upstream in the first and second quarters are smoothed out.  
Differences between quarters suggest a temporal structure
 which could be included if modelling observations from consecutive years. 
A wider range of IQA values for the first and forth quarter was detected in the exploratory analysis. 
Accommodating different scale parameters is not worthy for analysis of single year data and
can otherwise by considered, possibly with interactions with power plant effects. 
Such an addition to the model needs to be balanced against the usual 
numerical difficulties encountered when increasing of dimensionality of the random effects.
Possible workarounds such as quasi-likelihoods, 
MCMC algorithms, possibly under the Bayesian paradigm, or 
approximations such as the proposed by~\citet{Rue2009} need to 
be tailored for the beta random effects models.
Sensitivity to priors under the Bayesian approach 
 would be an issue for such model 
and might worsen with a larger numbers of random effects. 
The data clone proved helpful in eliminating effect of priors 
and assessing identifiability, at the expense of a greater computational effort.

\section{Conclusion}
A beta regression mixed model
including random effects associated with grouping units on a hierarchical model structure
is adopted for the analysis of 
two datasets with response variables on the unit interval, 
one on worker's life quality and another on water quality.
Different approaches were adopted for likelihood computations, 
and in particular we reported results for
numerical (Laplace) approximation and the sampling based 
algorithm of data cloning.
For the data analysis we use the strategy of fitting and selecting models using 
likelihood computations via the Laplace approximation followed by a detailed 
further assessment of the best model by data-cloning.

The first analysis shows the Brazilian industry life quality index is influenced 
by industry size and workers income 
with relevant random effects associated with the federative units. Findings based on the data analysis are compatible with subjective information validating social science's hypothesis.
For the second no negative effects of the damns on the water quality index was detect, 
which is relevant for licensing power plants operators. 
The beta random effects model accommodates environmental effects 
not fully captured by measured variables.
The random effects allows for a parsimonious model whilst 
considering extra sources of variation 
and a grouping structure.

Likelihood inference methods and algorithms were implemented using 
numerical approximations to integrate out the random effects from the likelihood computations.
 Confidence intervals based on profile likelihood were obtained with distinct results 
 from the ones obtained by asymptotic quadratic approximations in particular 
 for the parameters associated with the random effects. 
Implemented algorithms are made available\footnote{http://www.leg.ufpr.br/papercompanions/betamix}.
In general we obtained stable results in our analyses, however 
computational burden and accuracy of likelihood computations 
can be prohibitive
with increasing number of parameters associated with random effects. 

Numerical marginal likelihood computations were 
compared with another inference strategy based on a MCMC scheme for cloned data. 
The data clone algorithm is a relatively new and promising proposal 
with little programming burden at the cost of increasing computational effort, which can be partially 
alleviated by parallel or multicore computations for the several cloning numbers and chains.
A particularly attractive feature is the possibility of investigating identifiability,
which holds for both data analysis considered here. 
Point and interval parameter estimates based on data-clone are comparable with the ones obtained 
by Laplace approximations. 
Profiling likelihoods with data cloning requires further developments \citep{Ponciano2009}.

Bayesian analysis is frequently used for analysis of hierarchical models 
and computationally corresponds to the step of the data cloning algorithm 
with no replicates of the data. Sensibility analysis on prior choice
is relevant but attenuated by data-cloning.
Mixing of MCMC chains and identifiability remains relevant.
An attractive alternative is to run data-cloning combined with 
integrated nested Laplace approximations \citep{Rue2009}
which can be adjusted to deal with beta mixed models.
This can substantially reduce 
the computational burden by avoiding the more time demanding MCMC schemes,
carefully checking for the usage of improper priors,
if not completely avoiding them.

\paragraph{Acknowledgements} We thanks 
Milton Matos de Souza and Sonia Beraldi de Magalh\~{a}es  from 
\textit{Servi\c{c}o Social da Ind\'ustria (SESI)}
for the IQVT data.
We also thanks the Paran\'a Energy Company (COPEL) and
Nicole M. Brassac de Arruda from the 
\textit{Instituto de Tecnologia para o Desenvolvimento - LACTEC}
for the IQA data.

\bibliographystyle{dcu}
\bibliography{betamixlik}

\end{document}